            \documentclass[twocolumn,showpacs,pra,aps]{revtex4}
            
           \usepackage{graphics}
          \begin{document}
           \title{Resource Limited Theories and their
           Extensions}
          \author{Paul Benioff\\
           Physics Division, Argonne National Laboratory \\
           Argonne, IL 60439 \\
           e-mail: pbenioff@anl.gov}
           \date{\today}

          \begin{abstract}This work is based on the idea that extension
          of physical and mathematical theories to include the amount of
          space, time, momentum, and energy resources required to determine
          properties of systems may influence what is true in physics and
          mathematics at a foundational level. Background material, on the
          dependence of region or system sizes on both the resources required
          to study the regions or systems and the indirectness of the
          reality status of the systems, suggests that one associate to each
          amount, $r$, of resources a domain, $D_{r}$, a theory, $T_{r}$, and a
          language, $L_{r}$. $D_{r}$ is limited in that all statements in $D_{r}$
          require at most $r$ resources to verify or refute.  $T_{r}$ is limited
          in that any theorem of $T_{r}$ must be provable using at most
          $r$ resources. Also any theorem of $T_{r}$ must be true in $D_{r}$.
          $L_{r}$ is limited in that all expressions in $L_{r}$ require at
          most $r$ resources to create, display, and maintain. A
          partial ordering of the resources is used to describe minimal
          use of resources, a partial ordering of the $T_{r}$, and
          motion of an observer using resources to acquire
          knowledge. Reflection principles are used to push the
          effect of G\"{o}del's incompleteness theorem on consistency up
          in the partial ordering.  It is suggested that a coherent
          theory of physics and mathematics, or theory of everything,
          is a common extension of all the $T_{r}$.

          \end{abstract}
          \maketitle

          \section{Introduction}
          As is widely recognized, quantum mechanics
          and its generalizations, such as quantum field theory, is a highly
          successful theory. So far it has survived every
          experimental test. Yet in spite of this, nagging problems
          remain. The problem of measurement is one.  Although the
          use of decoherence to solve the problem \cite{Zurek,Zeh}
          helps in that it explains the existence of the pointer
          basis in measuring apparatuses, questions still remain
          \cite{Adler} that are related to whether quantum
          mechanics is really a theory of open systems only or
          whether there is a system such as the universe  that can
          be considered to be closed and isolated.  This is the approach
          taken by the Everett Wheeler interpretation \cite{Everett,Wheeler}.

          There are other more fundamental questions such as, why
          space-time is 3+1 dimensional, why there are four
          fundamental forces with the observed strengths, what
          the reason is for the observed elementary particle mass
          spectrum, and why the big bang occurred. Another
          basic question relates to why quantum mechanics is the correct
          physical theory. There are papers in the literature
          that address some of these questions by attempting to
          show that if things were different then life could not
          have evolved or some physical catastrophe would happen
          \cite{Tegmark,Vandam,Hogan,Barrow}. However these are all
          heuristic after-the-fact types of arguments and do not
          constitute proofs. The possibility of constructing a
          theory to explain these things, as a "Theory of
          Everything" or TOE represents a sought after goal of
          physics \cite{Barrow,Weinberg,Greene,TegmarkTOE,Schmidhuber}.

          Another very basic problem concerns the relation between
          physics and mathematics. The view taken by most
          physicists is that the physical universe and the
          properties of physical systems exist independent of and
          a-priori to an observers use of experiments to construct
          a theory of the physical universe. In particular it is
          felt that the properties of physical systems are
          independent of the basic properties of how an observer
          acquires knowledge and constructs a physical theory of
          the universe.  This view is expressed by such phrases as
          "discovering the properties of nature" and regarding
          physics as  "a voyage of discovery".

          A similar situation exists in mathematics. Most
          mathematicians  appear to implicitly accept the realist
          view. Mathematical objects have an independent,
          a priori existence independent of an observers knowledge of them
          \cite{Frankel,Shapiro}.  Progress in mathematics consists
          of discovering properties of these objects.

          This is perhaps the majority view, but it is not the
          only view.  Other concepts of existence include the
          formalist approach and various constructive approaches
          \cite{Heyting,Bishop,Beeson,Svozil}.  These approaches will not
          be used here as they do not seem to take sufficient
          account of limitations imposed by physics.  These include
          limitations resulting from the physical nature of language \cite{BenLP}.

          This realist view of physics and mathematics has some problems.
          This is especially the case for the widely accepted
          position that physical systems exist in and determine properties of a space-time
          framework.  However, mathematical objects exist outside of space-time and
          have nothing to do with space-time. If this is the case,
          then why should mathematics be relevant or useful
          at all to physics?  It is obvious that they are closely
          entwined as shown by extensive use of mathematics in theoretical
          physics, yet it is not clear how the two are related at
          a foundational level.

          This problem has been well known for a long time.  It
          was expressed by Wigner \cite{Wigner} in a paper entitled
          {\it The Unreasonable Effectiveness of Mathematics in the
          Natural Sciences}. A related question is, Why is
          Physics so Comprehensible? \cite{Davies}.

          Another foundational issue is based on the universal
          applicability of quantum mechanics.  It follows that all
          systems, including experimental equipment, computers,
          and intelligent systems are quantum systems in
          different states. The macroscopic aspect of these systems
          does not change their quantum mechanical nature.

          It follows that the process of validation (or
          refutation) of any theory, including quantum mechanics,
          is a quantum dynamical process described by quantum
          dynamical evolution laws. One sees then that quantum
          mechanics must in some sense describe its own
          validation by quantum systems.  However almost nothing
          is known so far about the details of such
          a description.

          These concerns form the background for this paper. This
          work begins with the observation that there is an aspect
          of physics that is faced daily by physicists, but is not included in
          physical or mathematical theories. This is the amount
          of physical resources, as space, time, momentum, and
          energy resources, required to carry out experiments and
          theoretical calculations. For experiments using large
          pieces of equipment and calculations requiring massive
          amounts of computing power, the resource requirements
          can be considerable.

          This use of resources  is not discussed in a
          theoretical context because of a strong belief that the amount
          of resources needed to carry out experiments and make theoretical
          calculations on different types of systems has nothing to do with
          the contents of physical theories being created and verified by this
          process. The material facts of what is true physically and properties
          of the theories making predictions supported by experiment, are believed
          to have nothing to do with the space time and energy momentum resources
          needed to do the experiments and carry out the computations. Extending
          this belief to a TOE  would mean that resource use by the knowledge
          acquisition process, whose goal is the construction of a
          coherent theory of mathematics and physics or TOE, has
          nothing to to with the contents of the TOE.

          The main purpose here is to take steps towards the possibility
          that this may not be correct, especially for foundational
          properties of physics and mathematics.  Included are
          questions regarding the strengths and existence of the
          four basic forces, why space-time is $3+1$ dimensional,
          the nature and reasons for the big bang and other
          general cosmological aspects, and why quantum mechanics
          is the correct physical theory.

          It should be strongly emphasized that the generally
          believed view of the independence between resource
          related aspects of carrying out experiments and
          calculations and the content of the theories
          created is true for the vast majority of physics and mathematics.
          There is ample evidence to support this view.  Probably the
          best evidence is that if it were not true, the
          dependence would have been discovered by now.

          However the fact that it is true for most systems and
          properties does not mean it is necessarily true for all. In
          particular, resource related aspects of doing experiments
          and calculations to create valid physical theories may
          influence the contents of the theories, at least at a
          very basic level.

          This work takes some initial steps to see if this
          possibility has merit.  The approach taken is an extension
          of the general ideas presented in \cite{BenTCTPM} and
          \cite{BenLP} and in references cited therein. The idea is to
          describe resource limited domains, theories and
          languages,  Each theory and domain is based on a limited amount of
          physical resources available to verify or refute the
          statements in the language. The relative strength of
          each theory depends on the amount of available
          resources.  Theories with more available resources are
          stronger than those with less.

          The next two sections give informal arguments that
          give some support to the possibility suggested here,
          that resource use may influence the basic contents of
          physical theories. The arguments are based on the
          relation between resource requirements and the size of
          the region or system being investigated. Another
          relation discussed is that between the indirectness of the reality
          status of systems and their size \cite{BenTCTPM}.

          These arguments lead to a description of resource limited
          theories, languages, and domains. This
          is provided in the subsections of Section \ref{RLT}.
          Included are a brief description of physical resources
          and a description of procedures, instructions,
          equipment, and purposes of equipment and procedures as
          components of the theory domains and languages. Other
          components include symbols strings as outputs of
          measurements and computations, and the implementation
          operation. These components are used to give descriptions
          of agreement between theory and experiment, and of theorem proofs
          in the theories (subsections \ref{AGE} and \ref{PTT}).
          Also the minimum resources required to determine the
          truth value of statements about properties of systems are
          discussed. The final subsection gives details on the effect of
          resource limitations on language expressions.

          Section \ref{POT} describes the use of the
          partial ordering of the physical resources to partially
          order the resource limited theories.  The
          following section describes briefly the dynamics of an
          observer using resources to acquire knowledge and
          develop physical and mathematical theories. The relation
          to the theories in the partial ordering is also noted.

          A characteristic of resource limited theories is that
          each theory includes parts of arithmetic and other theories.
          As such one expects G\"{o}del's incompleteness theorems
          \cite{Godel,Smullyan} to apply.  It is assumed that the
          resource limitations do not affect the validity of these
          theorems. One concludes from the second theorem that none
          of the theories can prove their own consistency, and that the
          same incompleteness applies to any extension proving the
          consistency of the first theory.

          It is possible to iterate the extension process
          and push the effect of G\"{o}del's theorem from theories
          with less available resources to theories with more
          available resources.  This is discussed in
          Section \ref{LRP} by the use of reflection principles
          \cite{Feferman,Fefermaninc} that are based on validity.
          Because of the resource limitations the reflection
          principles have to be applied separately to each individual
          sentence rather than to all sentences at once in a
          theory.

          Limit and consistency aspects of a TOE are discussed in
          Section \ref{PACTPM}.  The possibility that a coherent
          theory of physics and mathematics, or a
          TOE is a common extension of all the theories is
          noted as is a problem that consistency poses for a TOE.
          The final section summarizes the paper and points out
          the need for work on aspects not considered here.

          It must be emphasized that the goal of this paper is to describe
          some properties of resource limited theories, domains, and
          languages, and the motion of observers using resources
          to develop theories. As such this work is only a very small initial
          step in the approach to a coherent theory of physics and mathematics
          or a TOE. The material presented here {\it does not represent in any
          way} a completed TOE capable of verification or refutation. To
          achieve this many important aspects, not treated here, must be
          described. These include but are not limited to
          probability and information theory aspects, a much more detailed
          description of available physical resources including a
          description of resources {\it within each theory}, and specification of the
          axioms of the theories. Also many of the well known
          physical theories, such as quantum mechanics, general
          relativity, and possibly string theory, would have to be
          included in some form.

          \section{Resources and Region Size}\label{RRS}
          It is useful to begin by noting the
          relation between theories and the size of the  systems and
          regions to which the theories apply. For regions whose size is
          of the order of the Planck length, $\sim 10^{-33}$cm, string theory
          is used.  For Fermi sized regions, $\sim 10^{-13}$cm, the strong
          interaction is dominant with QCD the appropriate
          theory. For larger regions, $\sim 10^{-8}$cm up to
          thousands of cm in size, electromagnetic interactions are dominant
          with QED  the appropriate theory. Finally for cosmological
          sized regions, up to $10^{28}$cm in size, gravity is the
          dominant interaction with general relativity the appropriate theory.

          It is also well known that to investigate events in a
          region of size $r$, probes with momentum $\geq \hbar /r$
          and energy $\geq \hbar c/r$ must be used. The latter
          follows from the fact that the characteristic time
          associated with a region of size $r$ is given by the
          time, $r/c$, it takes light to cross the region.
          Here $\hbar$ is Planck's constant divided by $2\pi$ and
          $c$ is the velocity of light. This sets a lower limit on
          the energy momentum of a probe required to investigate
          events in regions of size $r$. It is a significant
          restriction for small $r$.

          What is, perhaps, not appreciated, but is well known by both
          theoretical and experimental physicists, is the fact
          that that there is another scale of physical resources
          associated with these regions of different sizes and
          their associated theories. These are the space time and
          energy momentum resources needed to carry out
          theoretical calculations and do experiments for the
          theories  and their systems relevant to regions of size $r$.

          The relationship between the size of the region investigated
          and the resources needed can be set out in general terms for
          both experiments and theory based computations. At present it
          appears impossible to do meaningful experiments and calculate
          the associated predicted outcomes for Planck sized objects as one
          does not know what to do or even if such objects exist. Because these
          objects are so small an extremely large or even infinite amount of
          resources are needed for such experiments and computations.

          To investigate Fermi sized objects, large accelerators and
          large amounts of energy are needed to produce the particle beams and
          maintain the relevant magnetic fields. Computations are resource
          intensive because the strong interaction makes
          a perturbation approach to QCD computations infeasible. The
          resources needed are large, but finite.
          Less resources are needed for relevant calculations and experiments
          on atomic and larger systems.  However more resources, in terms
          of very large telescopes, on and near earth, and long
          viewing times with very sensitive detectors, are needed to
          investigate cosmological sized objects, especially those
          that are very far away.

          The relations between resources needed and the size of
          the region investigated is shown schematically in
          Figure 1.   The ordinate shows a characteristic size parameter
          of the object being investigated. The upper limit shows the
          present age of the universe in cm and the lower limit is the Planck
          length in cm.

          The  first abcissa label, resource use, denotes the amount of resources
          required to carry out theoretical predictions and to do
          experiments on the object being investigated. The amounts increase
          from left to right as shown by the arrow. Values
          are not given because it is at present an open question how
          to quantify the resources required. Also for this paper
          there is no need to quantify the resources.

          \begin{figure}
          \begin{center}
           \resizebox{200pt}{200pt}{\includegraphics[214pt,525pt][445pt,710pt]{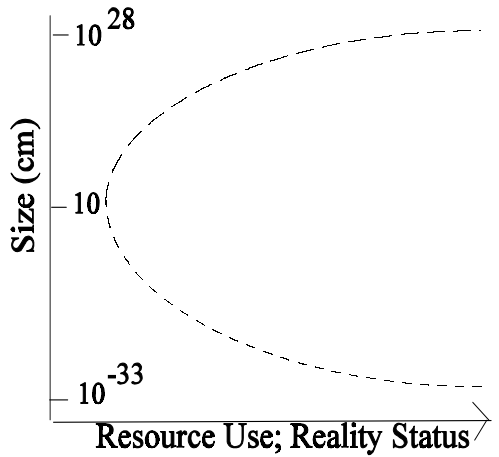}}
           \end{center}
          \caption{A schematic plot of the resource use and indirectness
          of reality status for systems of different sizes. Resource
          use refers to the amount of resources
          needed  to carry out calculations and experiments. Reality
          status is a measure of the number of layers of theory and
          experiment needed to give properties of systems. Additional
          details are in the text.}\label{size}
          \end{figure}

          The curve is a freehand or arbitrary schematic representation of
          the dependence of the resources required to carry out experiments and
          theoretical calculations on the characteristic size of
          the object being investigated. The arbitrariness, or lack of
          knowledge of curve details, is denoted by the
          dashes in the curve. In spite of this a curve, such as that in
          Fig. \ref{size}, is useful to represent some properties of the
          dependence that one does know. This is that the curve has
          two branches and that each branch must approach a limit. Here
          these limits are taken to be the Planck length and  $c$ times
          the age of the universe.  If one feels these limits are two restrictive
          they may be changed.  The important point is that there seem to be  such limits.

          The presence and location of the minimum
          represents systems whose size, real or perceived, is such
          that we can directly observe them. Many of these objects can be
          directly examined and handled to determine
          directly observable properties.  No experiments or theory is
          required as the properties can be determined directly by
          our senses.  Included are such properties as "this rock is
          heavy, hard, and brown","the horizon looks flat", "the sun is
          hot, bright, and moves through the sky".  The size of the sun is not
          the actual size but is the size perceived by us,
          which is a few cm.

          These directly perceived properties belong in the region of
          minimal resources required because  they are direct and
          uninterpreted. No theory or experiment is used to
          explain why anything happens or what its physical
          properties are. The resource location of the minimum of
          the curve is arbitrary. It is not set at $0$ resources to
          allow freedom in the choice of how resources are quantified.

          The ordinate location of the minimum represents sizes of
          objects that can be directly observed or perceived.  It is a
          broad minimum ranging over sizes of  the order of 1 cm to
          100 cm. To reflect this the minimum is arbitrarily
          set at 10 cm. The size range of the minimum is also representative
          of our size.  The reason is that our size is of the order of
          the (real or perceived) sizes of all systems that we can directly experience.

          \section{Size and Indirectness of the Reality Status}
          \label{SIRS}

         There is another quite different aspect of theories,
         theoretical calculations, and supporting or refuting experiments
         that is relevant to Figure 1. This is the
         indirectness of the reality status as a function of the size of
         physical systems  \cite{BenTCTPM}.

         To see this one notes that the validity of an experimental
         test of a theoretical prediction depends on the fact that each
          piece of equipment used in the experiment is properly
          functioning.  But the proper functioning of each piece of
          equipment depends in turn on other supporting theory
          and experiments which in turn $\cdots$. As an example suppose an
          experiment to test the validity of a theory at some point uses
          two pieces of equipment, $E_{1},E_{2}$. The validity of this
          experiment as a test depends on the proper functioning of $E_{1}$
          and $E_{2}$. However, the proper functioning of $E_{1}$ also
          depends on some theory which may or may not be the same
          as the one being tested, and also on some other experiments each
          of which depend on other pieces of equipment for their
          validity. This argument then applies also  to the experiments
          used to validate the theory on which the proper functioning of
          $E_{1}$ is based. Similar statements can be made for the
          proper functioning of $E_{2}$.

          Basic examples of such equipment are those that measure time and distance.
          The truth of the assertion that a specific system, called a clock,
          measures time depends on the theory and experiments
          needed to describe the functions of the clock components
          and the proper functioning of the clock components. The
          conclusion that a particular piece of equipment measures
          time depends on the conclusions that each component of the
          equipment functions properly. Similar
          arguments can be made for distance measuring equipment
          and equipment for measuring other physical parameters.

          Computations made to generate theoretical predictions
          have the same property. A computation is a sequence of
          different steps each performed  by one or more pieces
          of equipment such as a computer. Here the proper functioning
          of the computer depends on theory, which may or may not be
          the same as the one for which the computation is made, and on
          experiments that support the theory needed to assert
          that the computer does what it is supposed to do.

          These arguments show that the validity of an
          experiment or theoretical computation depends on a
          downward descending network of theories,
          computations, and experiments.  The descent terminates at the
          level of the direct, elementary observations that were
          discussed before. As was noted these require no theory
          or experiment as they are uninterpreted.

          The indirectness of the reality status of systems and
          their properties is measured by the depth of descent between
          the property statement of interest and the direct
          elementary, uninterpreted observations of an observer.
          This can be described approximately as the number of layers
          of theory and experiment between the statement of
          interest and elementary observations. The dependence on
          size arises because the descent depth, or number of
          intervening layers, is larger for very
          small and very large systems than it is for moderate
          sized systems.

          This line of argument gives additional support to the basic
          nature of the direct elementary observations perceived by
          an intelligent system. It is also shown by the curve in
          Figure 1 with the second abcissa label as a measure of the
          indirectness of the reality status of different sized objects.
          The indirectness can be roughly represented by the number of
          layers of theory and experiment between elementary observations
          and the theory calculations and experiments
          that are relevant for the object being investigated.

          The relation between the two abcissas suggests that
          resource use can be included by considering resource
          limited theories, domains, and languages and their
          relation to observers use of resources to develop
          theories.  Initial steps in this direction are carried
          out in the following sections.

          \section{Resource Limited Theories, Domains and Languages} \label{RLT}
          Before describing resource limited theories, domains, and
          languages, it is useful to give a brief description of
          physical resources.
          \subsection{Physical Resources}\label{PR}
          Here physical resources are considered to consist of space, time, momentum, and
          energy. If space and time is $d+1$ dimensional, then the amount,
          $r$, of resources available  is a $2d+2$ dimensional
          parameter $r_{1},r_{2}\cdots ,r_{2d+2}$. Each of the  parameters can
          be taken to be continuously varying or it can be considered to be
          discrete.\footnote{It is tempting to combine momentum and energy
          with space and time and let $r$ be
          a $d+1$ dimensional vector $r_{1},r_{2}\cdots ,r_{d+1}$
          where each $r_{j}$ denotes the available number of phase space cells
          for the $jth$ dimension, and $d$ is the number of space
          dimensions. The number of phase space cells of unit
          volume $\hbar^{d+1}$ associated with $r$ is given by
          $N_{r} = \prod_{j=1}^{d}r_{j}$. Here $\hbar$ is Planck's
          constant divided by $2\pi$. However this will not be
          done here.} Since the concerns of this paper are independent
          of which choice is made, the choice of a discrete or continuous $r$ will be
          left to future work.

          Each parameter, $r_{j}$, of the $2d+2$ parameter description of $r$ is a
          number indicating the amount of the $jth$ resource available. The $d$
          space parameters $r_{[1,d]}= \{r_{1},\cdots ,r_{d}\}$ and one time
          parameter $r_{d+1}$ give the amount of space and time
          available. Similarly the $d$ momentum parameters $r_{d+2},\cdots,
          r_{2d+1}$ and energy parameter $r_{2d+2}$ give the
          amount of momentum and energy available.

          Here it is also useful to consider a resource space
          whose elements are the $2d+2$ dimensional $r$.  The
          space has a partial ordering given by that defined for
          the resources.  That is $r\geq r^{\prime}$ if $r_{j}\geq
          r^{\prime}_{j}$ for all $j=1,\cdots ,2d+2$. This space
          represents a background for description of the resource
          limited theories and motion of observers developing
          theories.

          This description is sufficient for this paper even
          though it is quite superficial and brief. Additional
          details, including quantification and other aspects, are
          left to future work.

          \subsection{Basic Resource Limitations}\label{BRL}
          Let $T_{r},D_{r},L_{r}$ be a theory, domain, and language
          associated with each value of $r$. $L_{r}$ is the language used
          by $T_{r}$ and $D_{r}$ is the domain or universe of discourse
          for $T_{r}$. Here $r$ is the maximum amount of space, time,
          momentum, and energy resources available to $T_{r}$, $L_{r}$, $D_{r}$.
          This puts limitations on the $T_{r},L_{r},D_{r}$.

          A domain $D_{r}$ is limited by the requirement that at most
          $r$ resources are needed to determine the truth value of any
          statement $S$ in $D_{r}$. Let $r(S)$ be the resources needed
          to determine the truth value of $S$, i.e. to verify or refute $S$.
          If $S$ is in the domain $D_{r}$, then
          \begin{equation}r(S)\leq r.\label{reslim}\end{equation}
          If more than $r$ resources are needed to  verify or
          refute $S$, then $S$ is not in $D_{r}$.

          The statements $S$ can be quite general. Included are
          statements about properties of procedures, instructions,
          equipment, computers, and many other physical and
          mathematical objects. Since $S$ often includes statements
          about procedures used to determine properties or systems, there can be
          many statements $S$ for a given system and property,
          each based on a different procedure  and with a different value
          of $r(S)$. Similarly properties can be quite
          general.  Included are properties related to experimental
          tests of theories, purposes of procedures and
          instructions, existence of systems, etc.. The main point
          is that $D_{r}$ is limited to those $S$
          that satisfy Eq. \ref{reslim}.

          Here the value of $r(S)$ is considered relative to the basic
          uninterpreted directly perceived properties. Any
          resource value associated with these properties
          is the zero point.  Thus for each $S$ $r(S)$ includes
          the resources needed to construct all the equipment
          needed to verify or refute $S$.

          Note that $D_{r}$ is closed under negation  as $r(S)=r(\neg S)$ ($\neg$ means not).
          However $D_{r}$ is not closed under conjunction or
          disjunction. For conjunctions this follows from \begin{equation}
          r(S),r(T)\leq r(S\wedge T)\leq r(S)+r(T).\label{conjineq}\end{equation}
          One sees from this that it is possible that $S$ and $T$
          are such that $r(S)\leq r$ and $r(T)\leq r$ but $r(S\wedge T)>r.$
          In this case $S$ and $T$ are in $D_{r}$ but $S\wedge T$ is not. Note too that
          $r(S\wedge T)< r(S) +r(T)$ occurs if procedures
          for determining the truth values of $S$ and $T$ use
          some of the same equipment. Also $r(S\wedge T)$ should be such as to
          avoid double counting of resources used to construct equipment
          used in both procedures. The same arguments hold for
          disjunctions as \begin{equation} r(S),r(T)\leq r(S\vee T)\leq r(S)
          +r(T).\end{equation}

          The theories $T_{r}$ are limited by the requirement that
          proofs of all theorems of $T_{r}$ require at most $r$
          resources to implement. Thus $S$ is a theorem of $T_{r}$
          if a proof of $S$ can be done using at most $r$
          resources. If $S$ requires more than $r$ resources to
          prove, then $S$ is not a theorem of $T_{r}$.

          This limitation follows directly
          from the physical nature of language \cite{BenLP}. If the physical
          representation of expressions of $L_{r}$ corresponds to states of
          systems in $D_{r}$, which is the case assumed here, then
          the representation corresponds to a G\"{o}del map of the
          expressions into system states in $D_{r}$.  In this case the
          provability of a statement corresponds
          to a statement about properties of systems that are in
          $D_{r}$.  As such, the proof statements are subject to
          the limitations of Eq. \ref{reslim}.

          Another limitation on $T_{r}$ is that
          (assuming consistency) all theorems of $T_{r}$ must be
          true in $D_{r}$.  It follows from this and the first limitation
          that no statement can be a theorem of a consistent $T_{r}$
          if it is false in $D_{r}$, requires more than $r$
          resources to verify, or more than $r$ resources to
          prove.

          The language $L_{r}$ must satisfy a limitation based on the
          physical nature of language.  All expressions
          $X$ in $L_{r}$  as strings of symbols are limited by the
          requirement that they need at most $r$ resources to create, display,
          and maintain. This includes symbol strings, as strings of numerical digits
          (i.e. as names of numbers), which are used in all computations, quantum or
          classical, as outputs of measurements, and as
          instructions or programs for experimental or computation
          procedures. It is possible that there are expressions in $L_{r}$
          which are sentences but have no interpretation  as statements in
          $D_{r}$ because the interpretation does not satisfy Eq. \ref{reslim}.

          In this paper some major simplifying assumptions are made.
          One is that there is no discussion about how the
          resources and the limitations are described within the
          statements of $T_{r}$.  All resource discussions here are assumed
          to take place in the metatheory of the theories $T_{r}$.
          This puts off to future work removal of this assumption,
          which is clearly necessary.

          Another assumption is that probabilistic and information
          theoretic aspects are not included here.  It is clear
          that this assumption must be removed if quantum
          mechanics is to be included in any detail.  This is
          especially the case if the universal applicability of
          quantum mechanics is taken into account.

          A third assumption is that one specific physical
          representation of the symbols and expressions of $L_{r}$
          is assumed. Specific details are not given here as an
          abstract representation is sufficient.\footnote{Quantum
          mechanical examples of language symbols and expressions
          include lattices of potential wells
          containing ink molecules and products of spin projection
          eigenstates of spin systems also localized on a lattice.
          More details are given in \cite{BenTCTPM} and especially
          in \cite{BenLP}.}  It is clear, though, that there are many
          different physical representations of expressions, each
          with their own resource characteristics.

          \subsection{Contents of the Theories and Domains}
          \label{CTD}
          \subsubsection{Procedures, Instructions, Equipment}
          Included in the domains of the theories are processes or
          procedures, instruction strings, equipment, and statements about
          the function or purposes of procedures or equipment,
          and other physical and mathematical systems.
          Associated with a process or procedure $P$ is a set of
          instructions $I_{P}$ (as a symbol string)  for using several pieces
          of equipment.  Here $E_{P} = \{E_{1},\cdots ,E_{n}\}$ denotes the
          equipment used by $P$. $I_{P}$ may also include instructions for
          assembling the equipment in $E_{P}$ in specified locations and
          instructions on when to use it. In this case $E_{P}$ includes
          equipment to measure space and time.

          Procedures also contain branches. An example is
          the procedure $P$: ``Use $E_{3}$ to place $E_{2}$ 3 meters away from
          $E_{1}$. Activate $E_{1}$ and $E_{2}$.  Read outcome of using
          $E_{2}$, if outcome is $01101$ do $P_{1}$ if outcome is $11010$
          do $P_{2}$''. Here $P_{1}$ and $P_{2}$ are two other
          procedures that may or may not contain branches.

          There are no specific limits placed on pieces of
          equipment $E$.  $E$ can be as simple as clocks
          and measuring rods or as complex and massive and large
          as telescopes and particle accelerators. Of course,
          larger more complex equipment requires more resources to
          assemble, use, and maintain than does smaller, less complex
          equipment.

          It is important here to clearly separate
          {\it purposes} of both procedures $P$ and equipment $E$ from {\it use}
           of $P$ and $E$. $I_{P}$ should not say anything about
          what $P$ does or what any equipment used in $P$ does or why it is used.
          No theory is involved or needed to carry out $I_{P}$. $I_{P}$ represents
          instructions that can be followed by robots, automata, or other
          well trained implementers. Implementers, such as robots,
          must be able to follow instructions very well
          without knowing what anything is for.

          The example of a branching $P$ given above,
          violates this requirement by saying what $E_{3}$ does,
          ``Place $E_{2}$ 3 meters away from $E_{1}$". This was done both for
          illustrative purposes and as an aid to the reader. A
          proper description of $I_{P}$ would include
          instructions for how to use  $E_{3}$
          without saying anything about what $E_{3}$ is used for
          (space measurement). A possible way of saying this might be
          ``activate $E_{3}$, move $E_{1}$ until outcome $3$ shows on $E_{3}$".

          The same holds for the activation part of $P$. This
          denotes a procedure such as plugging cords into an electric
          socket. The implementer need not know that the
          procedure turns on $E_{1}$ and $E_{2}$ in order to
          follow the instructions. Activation may include observation of
          lights to determine if the equipment is on and
          properly functioning.

          The example $P$ also includes the component ``Read outcome on $E_{2}$,
          if outcome is $01101$ do $\cdots$".  This implies the
          direct reading of a symbol string showing in
          some part of $E_{2}$. No equipment is used as this is a
          direct uninterpreted observation.  No theory is used to
          make the observation and the implementer does not have to know
          whether the outcome is or is not a number
          or a symbol string to compare it with $01101$.\footnote{Note that the
          instructions either have to specify the ordering of
          reading the output symbols or a standard ordering must
          be assumed.  This is needed to convert the outcome
          $\{0,0,1,1,1,\}$, as an unordered collection of symbols,
          to the symbol string $01011$.} However the procedure may include
          instructions that are equivalent to using a piece of equipment
          $E_{4}$ to read $E_{2}$. This is useful in case it is difficult
          to read the output of $E_{2}$ and it is much easier to read the
          output of $E_{4}$ than of $E_{2}$.

          \subsubsection{Purposes}
          Associated with each procedure $P$, equipment
          $E$, and instruction string $I$, is a purpose
          $A$.  These denote what the procedure, piece of equipment,
          or instruction string does.  Examples of $A$  for procedures are
          ``prepares a system in state $\rho$ to $n$ figures",
          ``measures observable $O$ to $n$ figures", ``computes
          $Tr\rho O$ to $n$ figures", ``measures time to $n$ figures".
          For equipment, examples are ``is a telescope with operating
          parameters ---", and ``is an accelerator with
          operating parameters ---", and for instructions, examples are
          ``is instructions for using P", etc..  The reason for the
          accuracy phrase ``to $n$ figures" will be discussed
          later.

          The empty purpose,  ``has no purpose", is also included.
          This accounts for the fact that most processes do nothing
          meaningful, and most states of physical systems
          are not pieces of equipment that do anything meaningful.
          Also most symbol strings are not instruction strings or
          are instruction strings for meaningless procedures. For
          example making a pile of rocks in a road may have a purpose as a
          barricade but this is not relevant here.

          Purpose statements are used to associate purposes
          with procedures, equipment and instruction strings.
          The statement $F(P,A)$ means that ``$A$ is the purpose of
          $P$". If $A$ is ``measures time to $n$ figures", then $F(P,A)$ is
          the statement ``$P$ measures time to $n$ figures". Depending on
          what $P$ and $A$ are $F(P,A)$ may be true or false.
          In a similar fashion $F(E,A)$ and $F(I,A)$ are purpose
          statements for $E$ and $I$.

          Another type of useful purpose statement refers to the truth
          value of a statement $S$. If $B$ is the purpose
          statement ``determines the truth value of $S$'', then
          $F(Q,B)$ means ``$Q$ determines the truth value of
          $B$''.  Note that $S$ can be a statement $F(P,A)$. Then
          $F(Q,B)$ means ``$Q$ determines the truth value of
          $F(P,A)$''.

          This raises the question about the possibility of an
          infinite regress where $F(Q_{n+1},B_{n+1})$ says that
          $Q_{n+1}$ determines the truth value of $B_{n+1}$ and
          $B_{n+1}$ is the purpose ``determines the truth value
          of $F(Q_{n},B_{n})$''. There is indeed a regress but the
          regress is finite.  The reason is that the procedures,
          equipment and theory become more elementary as $n$
          increases with the regress terminating at the level of
          direct uninterpreted sense impressions of the type
          discussed in Section \ref{RRS}. The regress corresponds
          to a descent through a network of theories,
          computations, and experiments, Section \ref{SIRS}.

          This can be seen directly by noting that use of a
          procedure $P$ in some experiment requires that the
          purpose statement $F(P,A)$ be true.  Clearly any
          procedure $Q$ whose intended purpose is to determine if
          $F(P,A)$ is true or false must be more basic and direct,
          and depend on theories and experiments requiring less
          resources and interpretation than that for for the experiment
          using $P$. It is clear that this avoids circular situations where
          the validity of the purpose statement, $F(Q,B)$, with
          the purpose $B$ given by ``determines the truth value of
          $F(P,A)$'', depends on a theory whose validity is being
          tested by an experiment using $P$.

          \subsubsection{Outputs as Symbol Strings}\label{OSS}
          As the above shows, outputs as finite strings of symbols
          are an essential part of procedures.  Any measurement or
          calibration equipment used in a procedure generates
          output.  It is also worth noting that any
          output that is a string of $n$ digits, does not in general
          denote a number.  Instead it is an $n$ figure representation
          of a number.

          It is worthwhile to discuss this a bit especially in
          view of the resource limitations on the $T_{r}$.  The
          $4$ digit output binary string $1000$ corresponds
          to a natural number as it is a name for one.  However
          output in the binary form of $1\times 10^{11}$ does
          not correspond to a natural number.  Instead it is a one
          figure representation of some  range of numbers.
          However $1.000\times 10^{11}$ is a natural number
          (binary base and exponent) as it is equivalent to  $1000$.

          The situation is similar for output strings considered
          as rational numbers. For instance the $6$ digit binary output
          $101.011$, which is equivalent to $101011. \times 10^{-11}$,
          does not correspond to a specific rational number.
          Rather it corresponds to a $6$ figure representation of
          some range of rational numbers. The point is that if one
          assumes that an output string such as $101.011$ of some
          measurement is a rational number, then one is led to the
          conclusion that $101.011+\epsilon$, where $\epsilon$ is
          an arbitrarily small rational number,  is not the output
          of the measurement.  While this is literally true it can
          quite easily lead to wrong conclusions about the accuracy
          of the measurement, namely that the measurement is
          infinitely accurate.  Similar arguments  hold for real numbers
          in that no output digit string represents a real number\footnote{Of
          course mathematical analysis deals easily with single
          symbol representations of real numbers such as $\pi, e,
          \sqrt{2}$ and their properties.  But these are not
          outputs of measurements or equipment readings.}

          This description for the binary basis extends to any $k-ary$
          basis with $k\geq 2$. However, the possible values of $k$ are
          limited because there is a limit in how much information can be
          packed into a given space-time volume \cite{Lloyd}.

          The same limitations hold for purposes $A$ of
          procedures $P$.  If $P$ requires at most $r$ resources
          to carry out and $P$ represents a measurement of a
          continuously varying property, such as momentum, then the
          purpose statement $F(A,P)$ must include the property
          measured and the number of figures used to represent the
          outcome. If $P$ measures momentum, or prepares a system
          in some quantum state $\psi$, then $F(A,P)$ must say
          ``$P$ measures momentum to n figures" or ``$P$ prepares
          state $\psi$ to $n$ figures".  A procedure $P$
          that measures momentum  or prepares $\psi$ with no $n$ figure qualifier,
          would require an infinite amount of resources to
          implement.  Also the outputs of some
          of the equipment used in $P$, would have to be real
          numbers and require an infinite amount of resources to display.

          For measurements of discrete valued properties such as
          spin projections in quantum mechanics, the ``$n$ figure"
          qualifier can be dropped.  However this is the case only if
          $P$ does not also measure the continuously variable direction
          of the magnetic field serving as the axis of quantization.

          \subsubsection{Implementation}
          As described the procedures $P$ and their associated instructions
          $I_{P}$ do not include their own implementation. Also most $P$ and
          $I_{P}$ do not include instructions on when and where they
          are to be implemented.

          This is taken care of by use of an implementation operation $Im$.
          This operation refers to the actual carrying out of a procedure $P$
          by use of the instructions $I_{P}$. Implementation of $P$ also needs
          to specify when and where $P$ is to be done.  This is done by use of
          procedures $P_{s-t}$ that measure space and time to $n$ figures.
          The value of $n$ depends on the procedure used.

          $Im$ operates on pairs of procedures $P,P_{s-t}$ and on $d+1$ tuples
          $ \underline{x}$ of $n$ figure binary strings.  The result of actually
          implementing $P$  at a location and time given by $\underline{x}$, as
          determined by use of $P_{s-t}$, is denoted by $Im(P,P_{s-t},\underline{x})$.
          Since $P$ uses equipment, $I_{P}$ must describe how to set up the equipment
          and  how to use it to implement $P$. $Im(P,P_{s-t},\underline{x})$ then
          puts the equipment used in $P$ in some final state.

          Many procedures are measurements or computations. In this case
          the outcome as a string of digits corresponds to part of
          the final state of the equipment used.  Define $Ou$ to
          be the operation that picks out the output. In this case
          $Ou(Im(P,P_{s-t},\underline{x}))$ is the outcome digit
          string obtained by implementing $P$ at $\underline{x}$
          as determined by $P_{s-t}$.

          The implementation operation is quite separate from procedures
          $P$ and their instructions $I_{P}$.  This is the case even for $I_{P}$
          that state when and  where $P$ is to be carried out.
          Also $I_{P}$ often include instructions regarding
          relative spacing and delay timing of the various
          components. In this sense the $I_{P}$ are similar to
          construction and operating manuals accompanying disassembled
          equipment. Operating manuals can talk in great detail about
          using equipment or implementing procedures, but this
          is quite different from the actual use or implementation.

          It should be noted that physical resources must be used to
          carry out the implementation operation. For any procedure $P$
          resources are considered to be used at the space time point
          at which $Im$ is carried out. This includes the location of
          the space and time region needed to implement $P$ and the momentum and
          energy resources used to implement $P$ in the space time
          region so located.

          To see how this works let $P$ be a procedure whose
          purpose is denoted by $A$  and $E_{P} = E_{1}\cdots ,E_{n}$ be the
          equipment used by $P$. The truth of $F(P,A)$ namely,
          that $P$ does what it is supposed to do, depends on the
          truth of $F(E_{P},A_{P})=\wedge_{j=1}^{n}F(E_{j},A_{j})$
          where $A_{j}$ is the purpose of $E_{j}.$

          Let $Im(P,P_{s-t},x)$ be the result of
          implementing $P$ at $x$ by use of $P_{s-t}$. One requires
          at a minimum that $F(E_{P},A_{P})$ be true over the space
          and time region associated with the point $x$ at which
          $P$ is implemented. This would be  especially relevant
          for procedures whose implementation destroys some of the
          equipment used.

          Let $Q$ be a procedure whose purpose is
          to verify or refute $F(E_{P},A_{P})$. That is, $Q$ is a
          procedure to ensure that all the equipment in $E_{P}$
          is properly working. Implementation of
          $Q$ at $x^{\prime}$ gives outcome $1(0)$ if
          $F(E_{P},A_{P})$ is true (false) at $x^{\prime}$, or
          \begin{equation} Ou(Im(Q,P_{s-t},x^{\prime}))=1
          \Longrightarrow F(E_{P},A_{P}).\end{equation}  Since
          $x^{\prime}\neq x$ in general, physical theory
          (and equipment monitoring as part of $P$) is used
          to ensure the truth of $F(E_{P},A_{P})$ for the space
          time region occupied by the implementation of $P$ at
          $x$. The theory used includes basic aspects such as the
          homogeneity and isotropy of space and time, and
          predictions regarding a small influence of the environment on the
          equipment in $E_{P}$ in going from $x^{\prime}$ to $x$.

          \subsubsection{Agreement between Theory and
          Experiment}\label{AGE}
          The contents of the $T_{r}$  and $D_{r}$ described so far
          can be used to describe procedures that are tests of agreement
          between theory and experiment. Here only a very simple situation
          is considered in which one single experiment and one single
          theoretical computation is sufficient to test for agreement
          between theory and experiment.  Discussions of tests that
          require use of statistics and repeated experiments will be
          deferred to future work when probability concepts are introduced.

          The instructions $I_{P}$ include
          instructions for the use of three procedures.  Included
          are $P_{ex}$, whose purpose is to measure a property specified to
          $n$ figures on a system prepared in a state specified to $n$
          figures,\footnote{For astronomical systems, state preparation is
          not possible.} $P_{s-t}$ to measure space and time to $n$ figures,
          and $P_{th}$ to compute a number to $n$ figures. The measurement will
          also give an $n$ figure result. For simplicity the
          same value of $n$ is used for each procedure.

          The output symbol string, computed by $P_{th}$, is an $n$ figure
          representation of a numerical theoretical prediction  for
          the experiment.  As such it represents a theorem of the theory
          being tested where the theorem is adjusted to take
          account of the n figure specifications of the system
          state and property being measured and the output of the
          measurement.

          Let $A_{ex},A_{s-t},A_{th}$ denote $n$ figure purpose phrases for
          $P_{ex},P_{s-t},P_{th}$. $A_{ex}$ says ``measures to $n$ figures a
          property $Q$ specified to $n$ figures on a system in a state $\alpha$
          specified to $n$ figures". $A_{s-t}$ says ``measures space and
          time to $n$ figures", and $A_{th}$ says ``computes to $n$
          figures the theoretical value for the $n$ figure specification of
          property $Q$ measured on a system in the state $\alpha$ specified to
          $n$ figures".

          The statement of agreement between theory
          and experiment for these procedures  is the
          statement \begin{equation}Ag\equiv
           Ou(Im(P_{ex},P_{s-t},x_{ex}))=Ou(Im(P_{th},P_{s-t},x_{th})).
          \label{Ag}\end{equation}  $Ag$
          says that the outcome of implementing $P_{ex}$ at $x_{ex}$
          determined by use of $P_{s-t}$ equals the outcome of
          implementing $P_{th}$ at $x_{th}$ determined by use of $P_{s-t}$.

          The goal is to determine the truth value of
          $Ag$. The truth of $Ag$ is a necessary, but not
          sufficient, condition  for agreement between theory
          and experiment for the prediction that system in state
          $\alpha$ has property $Q$. The other necessary condition
          is that the three procedures have the purposes
          $A_{ex},A_{th},A_{s-t}$.  This is expressed by the requirement
          that the statement
          \begin{equation} Pur\equiv F(P_{ex},A_{ex})\wedge F(P_{s-t},A_{s-t})
          \wedge F(P_{th},A_{th})\label{Pur}\end{equation}
          must also be true. The truth of both $Ag$ and $Pur$ is
          necessary and sufficient for agreement between theory
          and experiment at $\alpha,Q$.

          The usual way of testing for agreement between theory
          and experiment is to actually implement the procedures
          as described here to determine if $Ag$ is true or false.
          This assumes the truth of $Pur$, which is based on
          other experiments and theory that agrees with
          experiment at other points.

          The well known use of resources to carry out experiments
          and theoretical computations is seen here by the
          requirement that resources are needed to verify or refute
          both $Ag$ and $Pur$. If $r(Ag)$ and $r(Pur)$ denote the
          resources needed, then $Ag$ and $Pur$ are in
          $D_{r}$ and $T_{r}$ if $r>r(Ag)$ and $r>r(Pur).$ The
          notion that $Pur$ and $Ag$ might also be theorems of
          some $T_{r}$, with resulting additional resource needs,
          is an intriguing but unexplored possibility.

          \subsubsection{Proofs of Theorems in $T_{r}$}
          \label{PTT} The contents of the $T_{r}$ can also be
          used to describe proofs of sentences in $L_{r}$. To see how
          this works, let $S$ be some statement such that $S$ is a theorem
          of $T_{r}$, or \begin{equation} T_{r}\vdash S.\label{trprv} \end{equation}
          This means that there exists a proof, $X$, of $S$ in $T_{r}$
          where $X$ is a string of formulas in $L_{r}$ such that each
          formula  in $X$ is either an axiom of $T_{r}$ or is
          obtained from some formula already in $X$ by use of a logical
          rule of deduction.

          With no resource limitations, which is the case usually
          considered, the process of determining if $T_{r}$ proves
          $S$ consists of an enumeration $X$ of theorems of
          $T_{r}$.  If $S$ is a theorem it will appear in $X$
          after a finite number of steps.  The proof $X$ with
          $S$ as a terminal formula will have a finite length.
          If $S$ is not a theorem it will never occur in an $X$
          and the process will never stop.

          Eq. \ref{trprv} is a statement in the metalanguage of
          the theories $T_{r}$. To give a corresponding statement
          in $L_{r}$ use is made of the physical representation of
          expressions in $L_{r}$. It was noted in subsection  \ref{BRL}
          that if  a physical representation of the expressions
          of $L_{r}$ is in $D_{r}$, then it corresponds to a G\"{o}del
          map $G$ of the expressions into states of systems in $D_{r}$.

          In this case theoremhood can be expressed using the contents of the
          $T_{r}$, Section \ref{CTD}. Let $P$ be a procedure
          acting on the states of physical systems described
          above. Let $\alpha$ be a state of some of the systems
          and $A_{\alpha}$ a purpose phrase in $D_{r}$ that says in effect
          ``repeatedly generate different states of the systems by
          a (specified) rule.  If and when state $\alpha$ appears on the
          designated subsystems, stop and output $1$".

          Let $B_{S}$, be a purpose phrase in the metalanguage that says
          ``enumerates proofs based on the axioms $Ax_{r}$ and stops with output
          $1$ whenever $S$ is produced at the end of a proof".
          Now require that $\alpha = G(S)$ and that $A_{\alpha}$
          satisfies \begin{equation} G(B_{S}) =A_{G(S)}.\label{gmprf}
          \end{equation} This requires $A_{\alpha}$ to be a
          physical purpose phrase that is equivalent under $G$
          to the purpose phrase for a proof enumeration until
          $S$ is generated.

          The statement that $P$ is a proof of $S$ of $T_{r}$ is given by
          the sentence $Y$ \begin{eqnarray}Y & \equiv &
          F(P,A_{G(S)})\wedge\nonumber \\ & \mbox{} & F(P_{s-t},A_{s-t})\wedge
          Ou(Imp(P,P_{s-t},x))=1. \label{Y}\end{eqnarray}
          Here $Ou(Imp(P,P_{s-t},x))=1$
          says that the output of implementing $P$ at $x$, based on
          use of $P_{s-t}$, is $1$.  This means the procedure stopped and $P$
          is a proof of $S$  under $G$. The sentences
          $F(P,A_{G(S)})$ and $F(P_{s-t},A_{s-t})$ are statements
          about the purposes of $P$ and $P_{s-t}$.

          Theoremhood for $S$ in $T_{r}$ is expressed by a
          sentence $Th_{r}(S)$ in $L_{r}$ saying that for all $x$ there
          exist procedures $P,P_{s-t}$ that satisfy $Y\equiv
          Y(P,P_{s-t},G(S),x):$\begin{equation}
          Th_{r}(G(S))\equiv\forall{x}\exists{P,P_{s-t}}Y(P,P_{s-t},G(S),x).
          \label{thrgs}\end{equation} If there is no such procedure
          then $S$ is not a theorem of $T_{r}$. Note that because
          the $T_{r}$ are incomplete, it does not follow from $S$
          not being a theorem that the negation of $S$ is a theorem.
          Each sentence is a theorem of $T_{r}$ if and only it can
          be proved with a procedure requiring less than $r$ resources to
          implement.

          Axioms play an important role in theories as
          they represent the input sentences for proofs.
          At this point it is not possible to specify
          the axioms, $Ax_{r}$, for each $T_{r}$. However some aspects
          are known. All $Ax_{r}$ consist of two components,
          the logical axioms and the nonlogical axioms. The logical
          axioms and logical rules of deduction are common to all theories
          as they represent a formal codification of the rules of thought  and
          logical deduction used to develop theories and to
          acquire knowledge. The nonlogical axioms distinguish the
          different theories as they should express exactly what a
          theory is about.

          Also all $Ax_{r}$ are limited by the requirement that
          each sentence in $Ax_{r}$ as a theorem of $T_{r}$ must
          satisfy the resource limitations on theorems of $T_{r}$
          stated earlier.  This has the consequence that for
          very small values of $r$ the $T_{r}$ are quite
          fragmentary as they contain very few sentences and even
          fewer as theorems. The resource limitations become
          less restrictive as $r$ becomes large.

          Subject to the above limitations all the $Ax_{r}$ would
          be expected to include axioms for arithmetic and axioms for
          operations on binary (or higher) names of numbers as
          $\tilde{0}-\tilde{1}$ symbol strings. This includes the use
          of these strings in expressions in $L_{r}$ corresponding to
          informal subscript and superscript labelling of variables,
          constants, functions and relations. Unary names are not used
          because arithmetic operations on these are not efficiently
          implementable \cite{BenRNQM}.

          The string axioms needed are those defining
          a concatenation operator, $\ast$, projection
          operators on different string elements, and string symbol
          change operators. Also included are two
          functions from strings to numbers denoting the length of a
          string and the number value of a string.

          It is expected that the $Ax_{r}$ will also include axioms
          for quantum mechanics and other physical theories.
          Further specification at this point is neither possible nor useful.
          The reason is that axioms and logical rules of deduction are in
          essence the initial conditions and dynamical rules for
          theorems of theories.  As such one wants to first
          investigate the theories in more detail to
          see what properties they  should have. This
          includes study of the dynamics of observers using
          resources to develop valid theories and inclusion of
          probabilistic and information theory aspects.
          Study of these and other aspects would be expected to
          give details on the specification of the $Ax_{r}$.

          \subsection{Minimal Use of Resources} \label{MUR}

          It is of interest to see in more detail how the basic resource
          limitations of subsection \ref{BRL} apply to the
          $T_{r}$. The main use of resources occurs through the
          implementation operation. This occurs because for any statement $S$
          the resources needed to verify or refute any statement $S$ are used by
          implementing the various procedures appropriate to $S$.
          This applies to all statements, including purpose
          statements, such as $F(P,A)$, provability statements, existence statements for
          different types of physical systems, and all others.

          A well known aspect of physics and other theories is
          that there are many different ways to prove something or
          to experimentally test some property of systems or to do
          things in general.  This is expressed here by procedure
          specific sentences such as those of Eqs. \ref{Ag}, \ref{Pur},
          and \ref{Y}.

          Let $S(\underline{P})$ be a procedure specific
          statement asserting that use of the procedures $ \underline{P}$
          shows that a specified system has a specified property.
          The underlined $ \underline{P}$ denotes possible use of more than one
          procedure. This is seen in the $Im$
          operation that operates on $2$ procedures and the $Pur$
          and $Ag$ statements based on $3$ procedures.

          Let $r(S,\underline{P})$ denote the resources needed to
          verify or refute $S(\underline{P})$. Since
          $r(S,\underline{P})$ is procedure dependent, there must
          be a set of procedures $ \underline{P}_{min}$ that
          minimizes $r(S,\underline{P})$. In this case
          \begin{equation}r(S,\underline{P}_{min}) =
          \min_{\underline{P}}r(S,\underline{P})\end{equation} is the least
          amount of resources needed to verify or refute a
          procedure specific statement $S(\underline{P})$.

          Let $S$ be the procedure independent statement asserting
          that a specified system has a specified property. Then
          $r(S,\underline{P}_{min})$ is also
          the least amount of resources needed to verify or refute
          $S$. Define $r(S)$ by \begin{equation}r(S)=
          r(S,\underline{P}_{min}). \label{rs}\end{equation} Here $r(S)$
          is the least amount of resources needed to verify or refute $S$.

          Note that one does not verify or refute $S$ by hunting
          through all possible procedures.  Instead one sets
          up procedures based on accumulated knowledge
          and resources spent. After a few tries one either
          succeeds in which case a procedure (or procedures) satisfying
          some $S(\underline{P})$ has been found. In this case the
          verification of $S$ follows immediately with no more
          resources needed.  If one fails then one either suspends
          judgement on the truth value of $S$ or concludes that it
          is false.

          This argument also holds for proof procedures. The well
          known recursive enumerability and non recursive nature
          of proofs shows up in the enumeration carried out by a
          proof procedure and not in trying lots of procedures.
          This is based on the observation that the resources needed to verify
          or refute $Th_{r}(G(S))$, Eq. \ref{thrgs}, are about\footnote{This
          allows for the small amount of additional resources needed
          to prove the quantified statement.} the same as are
          required to determine the truth value of
          $Y(P,P_{s-t},G(S),x)$, Eq. \ref{Y}, for the least resource
          intensive procedures.  The quantification
          over space time locations of the implementation operation is
          taken care of by including in the axioms the statements
          of homogeneity and isotropy of space and time. It
          follows from this that the resources required to verify
          or refute a statement are independent of where and when
          the appropriate procedures are implemented.

          The value of $r(S)$, Eq. \ref{rs}, represents the least
          value of $r$ for which the statement $S$ appears in $D_{r}$.
          All $D_{r}$ with $r\geq r(S)$ contain $S$, and
          $S$ is not in any $D_{r}$ where $r<r(S).$ In this sense
          $r(S)$ is the value of first appearance of $S$ in the $D_{r}$.
          The same argument holds for theorems. If $S$ is a theorem of
          $T_{r}$ then $r(S)$ is the $r$ value of first appearance of
          $S$ as a theorem in $T_{r}$.

          It is of interest to note that sentences $S$ that are theorems have two
          $r$ values of first appearance.  The first value, which
          is usually quite small, is the smallest $r$ value such
          that $S$, as a language expression, first appears in $L_{r}$.  The
          second much larger value is the value at which $S$ first
          becomes a theorem of $T_{r}$. If $S$ is not a theorem,
          then the second value is the value at which $S$ first
          appears in $D_{r}$.

          In a similar vein, the elementary particles of physics
          have resource values of first appearance in the $D_{r}$.
          To see this let $S$ be an existence statement
          for a particle type, such as a positron. Positrons exist
          only in those domains $D_{r}$ such that $r\geq r(S).$
          Statements regarding various properties of positrons
          also have $r$ values of first appearance. All these
          values are larger than $r(S)$.

          It should be noted that it is likely that there is no way to
          determine the values of $r(S)$ or $r$ values of first
          appearance of various properties.  Even if it were
          possible, one would have the additional problem of
          determining which procedure is most efficient.

          \subsection{Resource Limitations on Language Expressions} \label{RLLE}
          As was noted earlier the physical nature of language limits $T_{r}$ in
          that all expressions as strings of alphabet symbols in $L_{r}$ are
          limited to those requiring at most $r$ resources to create, display, and
          manipulate the expressions. This includes all symbol
          strings, as outputs and as formulas or words in $L_{r}$.

          To understand this better,  for each $a$ in the alphabet
          $\mathcal{A}$ of $L_{r}$, let $P_{a}$ be a procedure
          whose purpose is to create a physical system in some
          state  that represents the symbol $a$. An expression $X$
          of length $n=L(X)$ of symbols in $\mathcal{A}$
          can be considered a function
          $X:\{1,2,\cdots,n\}\rightarrow\mathcal{A}$. Let $p$ be
          an ordering rule for creating and reading $X$.  For
          instance $p$ can be a function from the natural numbers
          $1,2,\cdots,$ to intervals of space and time where
          $p(1)$ is the space and time interval between $X(1)$ and
          $X(2)$ and $\cdots$ $p(n-1)$ is the interval between
          $X(n-1)$ and $X(n)$. As such $p$ corresponds to a path along
          which the symbols of $X$ are created and displayed.
          Let $P_{X,p}$ be the procedure whose purpose is to use
          the $P_{a}$ to create $X$ according to $p$.

          The resources needed to implement $P_{X,p}$ depend on
          those needed to implement $P_{a}$ and to construct $X$
          according to rule $p$. Let $\Delta$ be the amount of
          physical resources used for each implementation of $P_{a}$.
          Here $\Delta = \Delta_{a}$ is assumed to be independent of
          $a$. It includes the amount of space and other resources
          needed to display a symbol.

          The amount of resources needed to create $X$ is given by
          $L(I(P_{X,p}))\Delta + r^{\prime}(P_{X,p}).$ the first
          part is the resources used by the instruction
          string or program for $I(P_{X,p})$ and the second part
          includes the resources needed to carry out $I(P_{X,p})$ or do
          $P_{X,p}$ and follow path $p$. It does not include the
          resources needed to display $X$.  These are given by
          $L(X)\Delta$.

          As states of physical systems, symbols created in a noisy
          environment require energy resources to maintain. If a symbol
          requires $\delta{E}$ energy resources per
          unit time interval to maintain, then maintaining an
          expression $X$ for m time intervals requires a total of
          $mL(X)\delta_{E}$ energy resources.  This assumes that none
          of the energy is recoverable.

          Putting the above together gives the result that the
          amount of resources needed to create, display, and
          maintain an expression $X$ for $m$ time intervals using
          instructions $I(P_{X,p})$  is given by
          \begin{eqnarray}
          r_{X,m,P_{X,p}}& = & L(I(P_{X,p}))\Delta  \nonumber \\
          & + & r^{\prime}(P_{X,p}) + L(X)\Delta +
          mL(X)\delta_{E}.\label{ResCDMS} \end{eqnarray}

          This equation denotes a $2d+2$ dimensional equation with
          one for each $i=1,2,\cdots,2d+2$. Each component
          equation is given by
          \begin{eqnarray}[r_{X,m,P_{X,p}}]_{i}& = &
          L(I(P_{X,p}))\Delta_{i} +
          [r^{\prime}(P_{X,p})]_{i}\nonumber \\ & + & L(X)\Delta_{i} +
          mL(X)\delta_{E}\delta_{i,2d+2}.\label{ResiCDMS}
          \end{eqnarray}Here the subscripts $i$ denote the $ith$
          component and $\delta_{i,2d+2} =1 (0)$ if $i= (\neq
          )2d+2$.

          Any theory $T_{r}$ with $r\geq r_{X,m,P_{X,p}}$ has $P_{X,p}$ in
          $D_{r}$. Also $X$ is in $L_{r}$. Here and in the following,
          unless otherwise stated,
          relations between two values of $r$ refer to all components of
          $r$.  However, if $[r]_{i}<[r_{X,m,P_{X,p}}]_{i}$
          for some $i$, then $X$ is not in $L_{r}$ as it requires
          too much of the $ith$ component of resources to create, display,
          and maintain.

          The previous discussion about minimal resources applies here
          in that there are many
          different procedures $P^{\prime} \mbox{ and instructions
          }I_{P^{\prime}}$, for creating symbols, and many different
          reading rules, $p^{\prime}$, and methods of maintaining $X$. The value of
          $r_{X,m,P^{\prime}_{X,p^{\prime}}}$ depends on all these
          parameters.  Also different physical systems in
          different states, from very large to very small, can be
          used to represent the alphabet of $L_{r}$.

          As before one is interested in the minimum value of
          $r_{X,m,P_{X,p}}$ for fixed $X$ and $m$ but varying $P$ and $p$.
          Finding a minimum for the $P$ and $I_{P}$ variations may be
          hard as this includes the algorithmic complexity of $X$
          \cite{Chaitin,MartinLof,Kolmogorov,Solomonoff}.
          However one would expect a minimal resource path $p$ to
          be a geodesic.  One also needs to account for variations
          in the extent and complexity of physical systems used to
          represent the alphabet symbols.

          For very small symbols quantum effects become important.
          This is especially the case if symbols are represented by
          coherent states of quantum systems.   Then the states
          must be protected against errors resulting from
          interactions with external fields and environmental
          systems. This is the basis for work on quantum error correcting
          procedures  for quantum computers.

          Here a fixed physical representation of alphabet symbols
          and a fixed path $p$ are assumed.  In this case Eq.
          \ref{ResCDMS} can be used to determine a number $N(r)$ that
          represents the maximum length of an expression $X$ whose creation,
          display, and maintenance for a time $r_{d+1}$ requires at most
          $r$ resources. To this end one replaces $L(I(P_{X,p}))$ by its
          approximate upper limit $L(X)$.  This accounts for the
          fact that, up to a constant, $L(I(P_{X,p}))$ is less than
          the length of a procedure that simply copies $X$. Also
          the $X$ dependence of $r^{\prime}(P_{X,p})$ is limited to a
          dependence  on $L(X)$ only.

          This allows one to define for each $i$ a
          number $N_{i}$ for any $r$ by \begin{equation} N_{i} =
          \max_{n}[n\Delta_{i}
          +[r^{\prime}(n,p)]_{i}+r_{d+1}n\delta_{E}\delta_{i,2d+2}\leq r_{i}].
          \end{equation} $N_{i}$ denotes the maximum length of any
          $X$ such that the $ith$ component of resources needed to
          create, display, and maintain $X$ is $\leq r_{i}$. Also
          $L(X)=n$. $N(r)$ is defined  by \begin{equation} N(r) =
          \min_{i=1,\cdots,2d+2}N_{i}.\label{Nr} \end{equation}
          $N(r)$ is determined by the most resource intensive
          component to create, display, and maintain an expression
          relative to the available resources.

          It should be noted that the resource limitations enter into
          $L_{r}$ and $T_{r}$ only through the requirement that the length
          $L(X)$ of all expressions in $L_{r}$ is less than some
          $N=N(r)$. One also sees that for moderate and larger
          values of $r$, the value of $N=N(r)$ for most physical
          representations of language expressions is very large.
          As such it is a  weak limitation especially when
          compared to the resources needed to determine the truth
          value of statements.

          \section{Partial Ordering of the $T_{r}$}
         \label{POT}

          The partial ordering of the resources $r=\{r_{1},\cdots
          ,r_{2d+2}\}$ can be used to partially order the theories $T_{r}$.
          In particular it is assumed here that $T_{r}\supseteq T_{r^{\prime}}$
          if $r\geq r^{\prime}$. Here $T_{r}\supseteq
          T_{r^{\prime}}$ means that the domain of $T_{r}$ includes
          that of $T_{r^{\prime}}$ and that $T_{r}$ is an extension of
          $T_{r^{\prime}}$ in that  every theorem of
          $T_{r^{\prime}}$ is a theorem of $T_{r}$. The latter is based
          on the observation that the resource limitations are weaker for
          $T_{r}$ than for $T_{r^{\prime}}$. As a result every proof $X$
          of a theorem in $T_{r^{\prime}}$ that does not include an axiom
          relating to resource limitations is a proof of the same theorem
          in $T_{r}$. Also axioms mentioning resource limitations have
          to be structured so that proofs including them do not
          generate contradictory theorems for different values of
          $r$. Whether this can be done or not is a problem for
          future work.

          If $r$ and $r^{\prime}$ are not in the domain of the partial
          ordering relation $\geq$, then the relation, if any, between
          $T_{r}$ and  $T_{r^{\prime}}$ is undetermined. This would be
          the case, for example, if $T_{r}$ has available twice the time
          resources and two thirds the space resources that are
          available to $T_{r^{\prime}}$.

            \begin{figure}
            \begin{center}
          \resizebox{200pt}{200pt}{\includegraphics[190pt,380pt][420pt,600pt]{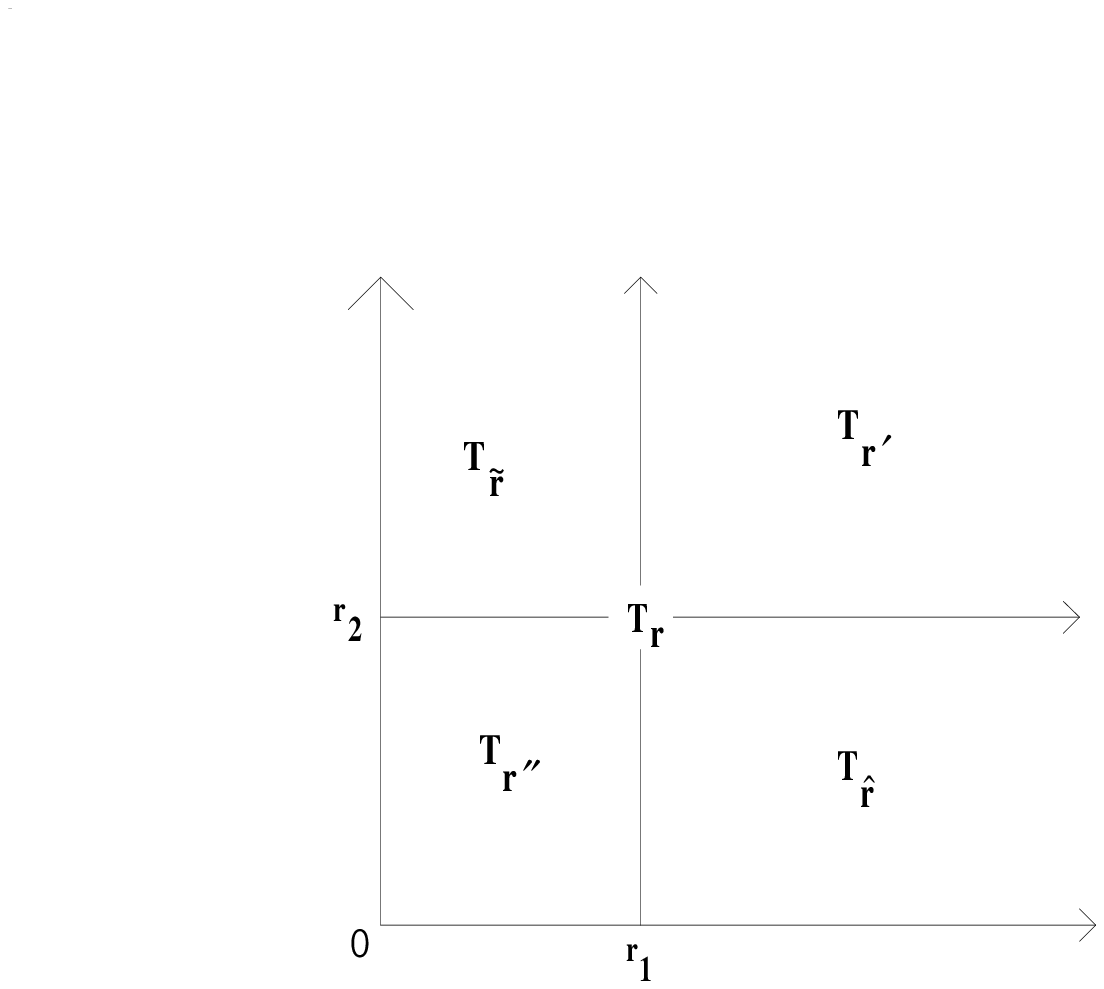}}
          \end{center}
          \caption{Partial Ordering of the Theories on a two
          Dimensional Resource Space. Theories in the upper right quadrant,
          such as $T_{r^{\prime}}$, are extensions or $T_{r}$.
          $T_{r}$ is an extension of theories in the lower left quadrant
          such as  $T_{r^{\prime\prime}}$. Theories in the other two
          quadrants are unrelated to $T_{r}$.}\label{Trrelations}
          \end{figure}

          These relations are shown in Figure \ref{Trrelations}
          where a two dimensional resource space is used to
          illustrate the relations. The figure coordinates show that
          the two resource components are $\geq 0$. The lines drawn through
          $T_{r}$ separate the theories into four
          quadrants.  The theories in the upper right quadrant,
          denoted by $T_{r^{\prime}}$, are all extensions of
          $T_{r}$, $T_{r}\subseteq T_{r^{\prime}}$.  $T_{r}$ is an
          extension of all theories in the lower left quadrant,
          such as $T_{r^{\prime\prime}}$, or $T_{r^{\prime\prime}}\subseteq
          T_{r}$. The theories in the upper left and lower right
          quadrants, such as $T_{\tilde{r}}$ and $T_{\hat{r}}$,
          are not related to $T_{r}$.

          The locations of various theories of physics and
          mathematics in the partial ordering are determined by the
          resource limitations on the domains, theories, and languages.
          This includes limitations based on resource use to prove
          statements, to determine the truth value of statements,
          and to limit the length of language expressions.

          One sees from this that expressions of a basic theory such
          as arithmetic are scattered throughout the $T_{r}$.
          There is no upper bound on the values of $r$ below which
          all arithmetic expressions are found.  It is also the case
          that for any $r$, no matter how large, almost
          all arithmetic expressions are found only in the $L_{r^{\prime}}$ where
          $r^{\prime}>r$. This holds even for the weak length
          limitation on expressions in the $L_{r}$. It is a consequence of
          the exponential dependence of the number of expressions on
          the expression length. The same holds for all names of the
          natural numbers as symbol strings in some basis.

          Many expressions of theories based on the real and complex numbers,
          such as real and complex analysis, quantum mechanics,
          QED, and QCD are also scattered throughout the $T_{r}$.
          However these are limited to expressions that contain at
          most variables and names of special mathematical
          objects such as $e,\pi,\sqrt{2},etc.$. These special objects are not
          random in that, for any $n$, they can be specified to $n$ figures by
          an instruction set $I_{P}$ as a symbol string of
          finite length that accepts $n$ as input
          \cite{Chaitin,MartinLof,Kolmogorov,Solomonoff}.  Almost all of
          the mathematical objects, such as real numbers, complex
          numbers, functions, states, operators, etc., are random.  Names for
          all of these cannot be found in any $L_{r}$ no matter how
          large $r$ is.

          It follows that almost all sentences  $S$ in these theories are
          infinitely long.  These expressions are in the limit
          language, $L_{\infty}$, only. They are not in $L_{r}$ for
          any finite $r$.

          Another way to state this is that quantum mechanics and many other other theories
          are limit theories. Each is a theory of first appearance for
          the parts of all the $T_{r}$ that are expressions and
          theorems for the theory being considered. This holds even for
          arithmetic whose expressions, including names, are of
          finite but unbounded length.

          \section{Resource Use by Observers}\label{RUO}
          The resource space and the  $T_{r}$,
          Figure \ref{Trrelations}, represent a background on
          which an intelligent system (or systems) moves in developing physical and
          mathematical theories and, hopefully, a coherent theory of
          physics and mathematics or a TOE. The main goal of
          interest for an observer (assumed equivalent to an
          intelligent system)  or community of observers is to
          develop physical and mathematical theories that explain
          their observations.

          Here the need for observers to use physical resources to
          acquire this knowledge is emphasized.  Observers start
          with elementary sense impressions and acts,
          uninterpreted by any theory, Sections \ref{RRS}
          and \ref{SIRS}.  They use physical resources to carry
          out experiments and theoretical calculations to develop
          physical theories that explain their impressions and
          suggest new ways to test the theories. Which resources
          an observer uses and what the resources are spent on are
          determined by the specific observer.  It depends on
          choices made and the goal of the process for each
          observer.

          It is clear from this that the process of using
          resources to develop a theory or theories to explain
          observations and results of experiments is a dynamical
          process.  To this end let $p$ denote a path in resource
          space that represents the resources used by  an observer
          where $p(t)$ denotes the total amount
          of resources used up to time $t$ by an observer.  If
          $dp_{i}(t)/dt$ is the time rate of change of the use of the $ith$
          component of the resources then \begin{equation}
          p_{i}(t)=\int_{0}^{t}\frac{dp_{i}(t^{\prime})}{dt^{\prime}}dt^{\prime}
          \end{equation} gives the time development of the use of the $ith$ resource
          component.
            \begin{figure}
          \begin{center}
          \resizebox{180pt}{180pt}{\includegraphics[190pt,380pt][420pt,600pt]{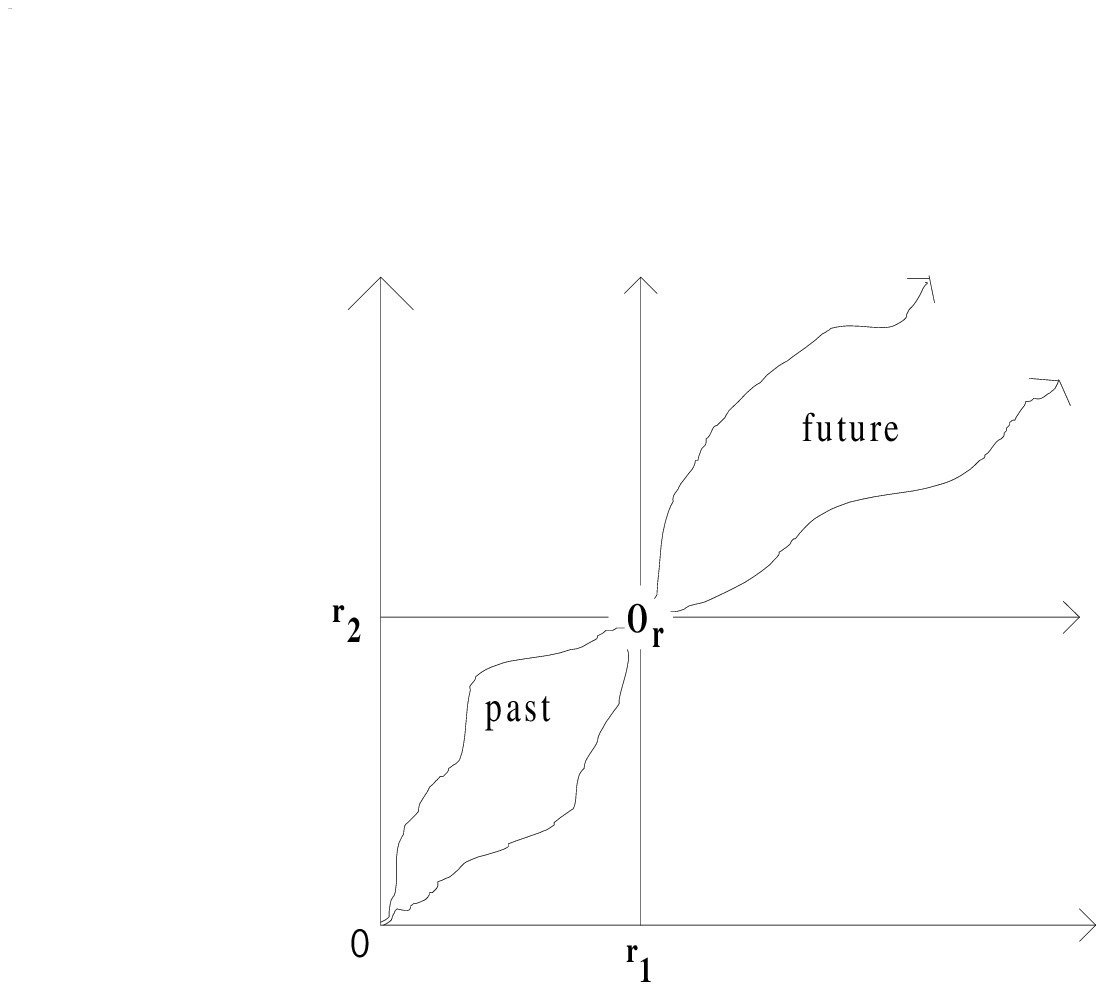}}
          \end{center}
          \caption{Two Paths Showing Use of Physical Resources by an Observer. $O_{r}$ shows
          the position of an observer after spending $r =p(t)$ resources by time $t$. Paths in the
          lower left and upper right quadrants, denoted as past and future, show use of
          resources at times before and after $t$. Also path gradients must be $\geq 0$
          everywhere.}\label{Obs}
          \end{figure}
           The motion of an observer using resources can be
          shown on a figure similar to Fig. \ref{Trrelations}.
          This is done in Figure \ref{Obs} which shows the
          location of an observer after having used $r=p(t)$
          resources at some time $t$. As was done for Fig.
          \ref{Trrelations}, $r$ is taken to be $2$ dimensional.
          The figure shows two out of many possible paths in the
          resource space that an observer can follow. available
          for an observer. The path gradients, $dp_{i}(t)/dt$, are
          $\geq 0$ everywhere.  This follows from
          the requirement that used resources cannot be
          recovered. Resources used before time $t$ are in the
          lower left quadrant, labelled as past, and resources
          used after time $t$ are in the upper right quadrant,
          labelled as future.

          The knowledge gained by an observer $O$ at time $t$ after using $p(t)$
          resources can be represented by a statement
          $\underline{S}_{p(t)}=\wedge_{j=1}^{n}S_{j}$ that is the
          conjunction of all statements verified or refuted by $O$
          after using $r=p(t)$ resources. That is \begin{equation}
          r(\underline{S}_{p(t)})=p(t).\label{restime}\end{equation}
          $\underline{S}_{p(t)}$ can include many types of component
          statements such as those about tests of agreement between theory and
          experiment. Also some or all of the component statements can
          be theorems. The number $n$ of statements depends on many
          things including what procedures an observer decides to
          implement in acquiring knowledge.

          As was noted before, Eq. \ref{conjineq}, the amount of
          resources spent to verify or refute
          $\underline{S}_{p(t)}$ can be less than the sum of the
          resources needed to determine the truth value of each
          component $S_{j}$ considered by itself.  The amount of
          resources spent to verify or refute $S_{j}$ in the conjunction is given
          here by $p(t_{j})-p(t_{j-1})$ where $t_{j}$ is the time
          resource used by $O$ to verify or refute $\underline{S}_{p(t_{j})}=
          \wedge_{k=1}^{j}S_{j}.$

          The connection between $O_{r}$ in Fig. \ref{Obs} and
          $T_{r}$ in Fig. \ref{Trrelations} follows from Eq.
          \ref{restime}.  In particular $\underline{S}_{p(t)}$ is
          a statement in $D_{p(t)}$ and in $T_{p(t)}$. It is also the case
          that each component statement $S_{j}$ in  $\underline{S}_{p(t)}$
          is in $D_{\Delta_{j}}$ and in $T_{\Delta_{j}}$ where $\Delta_{j} =
          p(t_{j})-p(t_{j-1}).$ Also any $S_{j}$ that is a theorem
          is a theorem of $T_{\Delta_{j}}.$

          It follows from Eq. \ref{conjineq} that all component
          sentences of $\underline{S}_{p(t)}$ are included in
          $D_{r},L_{r}$, and $T_{r}$ where $r=p(t).$ $T_{r}$ should prove some
          of the verified sentences in $\underline{S}_{p(t)}$ and prove
          none of the refuted sentences. Also $T_{r}$ and $D_{r}$
          contain many other sentences obtained by observers
          following different resource use paths and
          choosing a different collection of statements to verify
          or refute. For instance if
          $\underline{S^{\prime}}_{p^{\prime}(t^{\prime})}
          = \wedge_{j=1}^{m}S^{\prime}_{i}$ is verified or refuted by
          following a different resource path $p^{\prime}$, then
          $\underline{S^{\prime}}_{p^{\prime}(t^{\prime})}$ and
          each component statement is in $T_{r}$ and $D_{r}$
          provided that $p^{\prime}(t^{\prime})=r.$

          \section{Local Reflection Principles} \label{LRP}
          As is well known, the goal of any theory, including the
          $T_{r}$, is to determine the truth value of statements.
          The only method available for a theory to determine truth
          values is by proof of theorems. However this  works if
          and only if the theory is consistent. All statements of
          inconsistent theories are theorems so there is no connection
          between theoremhood and truth or falseness.

          This also applies to the partially ordered $T_{r}$. For this reason,
          it would be desirable if the $T_{r}$ could prove
          their own consistency or validity.  However, this is not
          possible for any theory, such as the $T_{r}$, containing
          some arithmetic \cite{Godel,Smullyan}. The same limitation applies also
          to any stronger theory that proves the consistency of the
          original theory. It is assumed here that the resource
          limited $T_{r}$ have the same properties regarding consistency
          proofs as theories with no resource limitations.

          Here reflection principles,
          based on validity statements  \cite{Feferman,Fefermaninc}, are
          used with the $T_{r}$ to push validity proofs
          up in the partial ordering of the $T_{r}$. In this way
          theories higher up in the ordering can prove the validity
          of theories lower down. To this end let $S$ be some statement such that
          $T_{r}$ proves $S$, Eq. \ref{trprv}.  Then $Th_{r}(G(S))$, given by
          Eq. \ref{thrgs} is a theorem of $T_{r}$.  This is expressed by
          $T_{r}\vdash Th_{r}(G(S))$, which says that the sentence $Th_{r}(G(S))$
          is a theorem of $T_{r}$, or that $T_{r}$ proves that it proves $S$.

          The validity of $T_{r}$ at $S$ is expressed by \begin{equation}
          Val_{r}(G(S))\equiv (Th_{r}(G(S))\Longrightarrow S).\label{valrs}
          \end{equation} $Val_{r}(G(S))$ is a sentence in $L_{r}$ which
          can be interpreted through $G$ to say that if $T_{r}$ proves
          $S$ is a theorem, then $S$ is true. Here one is using Tarski's
          notation that assertion of a statement $S$ is equivalent to the
          truth of $S$ \cite{Tarski}. This means that if $Val_{r}(G(S))$
          were a theorem of $T_{r}$, then one could conclude from $T_{r}\vdash
          Th_{r}(G(S))$ that $T_{r}$ proves the truth of $S$.

          The problem is that because $T_{r}$ cannot prove its
          own consistency it cannot prove validity statements such
          as $Val_{r}(G(S))$. Reflection principles
          \cite{Feferman,Fefermaninc} are used here to extend the $T_{r}$
          with validity statements for the sentences
          in $T_{r}$.  Because of resource limitations, the extensions must
          be considered separately for each $S$ rather than
          adding validity statements for all sentences of $T_{r}$
          to the axioms of $T_{r}$ \cite{Feferman,Fefermaninc}.
          Also since the axiom sets $Ax_{r}$ are not specified
          in any detail, the addition is taken care of by requiring that
          the axiom sets $Ax_{r}$ are such that theories higher up
          in the partial ordering can prove the validity of theories
          lower down.

          In this case the $T_{r}$ have the property that for each $S$ for
          which Eq. \ref{thrgs} holds, there exists a
          theory $T_{r^{\prime}}$ with $r^{\prime}>r$ that proves
          the validity of $T_{r}$ at $S$ or \begin{equation}T_{r^{\prime}}\vdash
          Val_{r}(G(S)).\end{equation} Since  $r^{\prime}>r$ implies
          that \begin{equation}T_{r^{\prime}}\vdash Th_{r}(G(S)),\end{equation} one has that
          $T_{r^{\prime}}\vdash S$. In this way $T_{r^{\prime}}$ reflects the
          validity of $T_{r}$ and proves that $S$ is
          true.\footnote{One cannot conclude directly from Eq. \ref{trprv}
          that $S$ is true because $T_{r}$ lacks a proof of its
          own validity at $S$.}

          This transfers the validity problem to $T_{r^{\prime}}$.  In order to
          conclude that $S$ is true, one needs to prove that
          $T_{r^{\prime}}$ is valid at $Th_{r}(G(S))$ and at
          $Val_{r}(G(S))$. This leads to an iterated application of
          the reflection principles generating a sequence of
          theories $T_{r_{n}}$ where $r_{n+1}>r_{n}$ and
          $T_{r_{n+1}}$ proves the validity of the relevant statements
          for $T_{r_{n}}.$ Based on G\"{o}dels
          second incompleteness theorem \cite{Godel,Smullyan} the
          iteration process does not terminate. Here this leads
          to  limit theories that have the same problem.  The limit
          theories are the usual theories with no
          bounds on the available resources.\footnote{Iteration of this
          process into the transfinite by use of constructive ordinals
          \cite{Feferman,Turing} and closure by the use of self truth
          axioms is discussed in the literature \cite{Fefermaninc}.}

          \section{Possible Approach to a Coherent Theory of Physics
          and Mathematics}\label{PACTPM}
          At this point little can be said about the details of a
          coherent theory of mathematics and physics or a TOE.
          However there are some properties of a TOE that
          would be expected if the partial ordering of theories
          and resource used by observers described
          here has merit.  These are the relation of a coherent theory
          to the $T_{r}$ and the problem of consistency.

          \subsection{Limit Aspects}\label{LA}
         As was seen in section \ref{POT} the expressions of
         arithmetic and other theories of physics and mathematics are
         scattered throughout the $T_{r}$ with the number of
         expressions and sentences first appearing in $T_{r}$
         increasing exponentially with the value of $r$.  This holds for
         arithmetic sentences and sentences of other theories with
         names of objects that are not random. However since names of most
         mathematical objects are infinitely long, so are
         sentences that include these names.

         As was noted earlier, it follows from this that theories
         of physics and mathematics with no resource limitations
         are limit theories or theories of first appearance of all
         the expressions appropriate to the theory being
         considered. Arithmetic is the theory of first appearance
         of all the arithmetic expressions of the $T_{r}$. Quantum
         mechanics is the theory of first appearance of all
         expressions in the parts of the $T_{r}$ that deal with
         quantum mechanics.  The same holds for other theories.
         They are all limit theories or theories of first
         appearance of the relevant parts of the $T_{r}$.

         If one follows this line of thought, then a coherent
         theory of mathematics and physics or TOE would also be a
         limit theory with expressions scattered throughout the
         partial ordering. In this case one would expect the TOE to be a
         common extension of all the $T_{r}$ rather than of just parts of
         each $T_{r}$.  In this case one expects that
         \begin{equation} T_{r} \subset TOE \label{TOE}\end{equation}
         holds
         for each $r$. That is any statement that is a theorem in
         some $T_{r}$ is also a theorem in TOE.  This requires
         careful inclusion of the resource limitations into the
         $T_{r}$ and the $Ax_{r}$ so that some obvious, and not so
         obvious, contradictory statements do not become theorems.
         Whether or not the TOE satisfies this condition has to
         await future work.

         \subsection{Consistency and a Coherent
         Theory}\label{CCT} Consistency poses a problem for a
         coherent theory of physics and mathematics or a TOE to
         the extent that this theory is assumed to really be a final theory
         \cite{Weinberg} in that it has no extensions. It was seen that
         G\"{o}del's incompleteness theorem on consistency\cite{Godel,Smullyan} and the
         use of reflection principles \cite{Feferman,Fefermaninc}
         push the consistency problem up the network but never get
         rid of it.  Also it follows directly from Eq. \ref{TOE}
         (and from the fact that a TOE includes arithmetic) that a
         TOE cannot prove its own consistency.

         This is problematic if a TOE is a final theory because if
         one extends a TOE to a theory proving that the TOE is
         consistent then a TOE is not a theory of everything.  It
         is a theory of almost everything. And the same problem
         holds for the extension.

         This situation is unsatisfactory.  However it is no worse than the
         existing situation regarding other theories such as arithmetic,
         quantum mechanics, and many other physical and mathematical theories.
         Each of these theories can express their own consistency, so none of
         them can prove their own consistency \cite{Godel,Smullyan}.  Such proofs
         must come from stronger theories which then have the same problem.
         Of course, there is no reason to doubt the consistency of these
         theories, and their usefulness shows that they are almost
         certainly consistent.

         For a limit or final theory \cite{Weinberg} one would like to do
         better and not leave the problem hanging. One solution might be to
         solve the problem axiomatically by including an axiom that asserts
         the existence of a consistent coherent theory of physics and
         mathematics. How the axiom is stated, such as whether or not it
         is in essence the strong anthropic principle
         \cite{Hogan,Barrow,BenTCTPM}, and the usefulness of this approach,
         will be left to future work.

         \section{Summary and Future Work}
         \subsection{Summary}
         A partial ordering of resource limited theories and their
         extensions has been studied as a possible approach to a
         coherent theory of physics and mathematics. Each theory
         $T_{r}$, domain $D_{r}$, and language $L_{r}$ has a limited amount $r$ of space,
         time, momentum, and energy resources available.

         The  resource limitations on the $D_{r}$ restrict all
         statements $S$ in $D_{r}$ to require at most $r$ resources
         to verify or refute. The statements can refer to
         processes, physical systems,  purposes of processes,
         implementations of procedures, and outcomes of experiments
         and whether they agree or disagree with theoretical predictions.

         Resource limitations on the $T_{r}$ require that all
         theorems are provable using at most $r$ resources. Also
         if $T_{r}$ is consistent, then all theorems of $T_{r}$
         must be true in $D_{r}$.

         A less restrictive limitation is that the language $L_{r}$ is
         limited to expressions, as strings of symbols from
         some alphabet, that require less than $r$ resources to
         create, display, and maintain.  This is expressed here by
         a length limitation on the expressions, given by Eq.
         \ref{Nr}, that is based on the essential physical nature of
         language \cite{BenLP}.

         The contents of the theories are described in some
         detail.  Included are procedures, equipment, instructions
         for procedures and purposes.  The implementation
         operation and its role in the use of resources is
         discussed. These components were used to give statements
         in $L_{r}$ that express agreement between
         theory and experiment, and provability of a statement
         $S$.  The role of G\"{o}del maps based on the physical
         nature of language in the provability statement was
         noted.

         It was noted that there are many different procedures
         for determining the truth value of a statement $S$. As a result
         there is a minimum amount $r(S)$
         of physical resources associated with determining the
         truth value of  $S$.  Based on this $r(S)$ is also the resource
         value of first appearance of $S$ in the $D_{r}$ and
         $T_{r}$. If $S$ refers to the existence of some
         elementary particle of physics then the particle first
         appears in $T_{r(S)}$ and in $D_{r(S)}$.

         A partial ordering of the theories is  based on the
         partial ordering of the resources $r$.  $T_{r^{\prime}}$
         is an extension of $T_{r}$ (all theorems of $T_{r}$ are
         theorems of $T_{r^{\prime}}$) if $r^{\prime}\geq r$, i.e.,
         if for all components $r_{i}$ of $r$, $r^{\prime}_{i}\geq
         r_{i}$. This requirement is a nontrivial condition that
         the axioms $Ax_{r}$ of each $T_{r}$ must satisfy. This is
         in addition to the requirement that no statement requiring $>r$
         resources to verify or refute can be a theorem of $T_{r}$.
         Also no false statement in $D_{r}$ can be a theorem of $T_{r}$.

         The motion of an observer using resources to develop
         theories was briefly discussed. It was noted that the
         amount $r$ of resources used by an observer can be divided
         into parts with each part being the resources used to
         verify or refute a statement. The collection of all
         statements verified or refuted by an observer, following
         some path $p$ of resource use, represents the total
         knowledge of the observer regarding development of
         physical and mathematical theories.

         A brief discussion was given of the use of reflection
         principles \cite{Feferman,Fefermaninc}
         to push the effect of G\"{o}del's second incompleteness theorem
         \cite{Godel,Smullyan} on the $T_{r}$ up in the partial ordering.
         This was done by the use of validity statements
         $Val_{r}(G(S)\equiv Th_{r}(G(S)\Longrightarrow S$ which state
         that $T_{r}$ is valid for $S$. Here it is assumed that the
         axioms $Ax_{r}$ are such that for each $S$ there is an $r^{\prime}>r$
         such that  both $Val_{r}G(S))$ and $Th_{r}(G(S))$ are theorems of
         $T_{r^{\prime}}$. G\"{o}del's theorem, applied to $T_{r^{\prime}}$
         leads to iteration of this process to
         limit theories with no bounds on the available resources.

         The possible use of the partial ordering of the $T_{r}$ as an
         approach to a coherent theory of physics and mathematics, or TOE, was
         briefly discussed.  It was noted that a TOE must be a
         limit theory that includes all the $T_{r}$, i.e. $T_{r}\subset TOE$.
         In this way a TOE includes arithmetic, quantum mechanics
         and other physical and mathematical theories, which are
         also parts of the $T_{r}$. This introduces a problem
         for consistency. Since a TOE can express its own
         consistency, it cannot prove its own consistency.  However
         if a TOE is a final theory with no extension, then the consistency
         problem for a TOE is left hanging.

         \subsection{Future Work}
         As the above suggests there is much to do. Probably the
         most important need is to extend the theories to include
         probability and information  theory concepts.
         It is expected that this will be important relative to
         observers spending resources to acquire knowledge and
         move towards a limit theory.

         Another basic need is to develop the description of the
         theories $T_{r}$ so that they describe the use of resources
         and the effects of limited availability of resources.  This is clearly
         necessary if the axioms of $T_{r}$ are to be such that no statement
         requiring more than $r$ resources to verify or refute is
         a theorem of $T_{r}$.

         The conditions imposed on the axioms $Ax_{r}$ in this
         work are quite complex.  At this point it is open if
         there even exist axiom sets that can satisfy all the
         conditions.  This needs to be investigated.

         Another assumption that must be removed is
         embodied in the use of Eq. \ref{Nr} to limit the length
         of language expressions.  The theories $T_{r}$ must take
         account of the observation that physical representations
         of language symbols and expressions as symbol strings can
         vary widely in size and resource requirements to create,
         display, maintain, and manipulate. There is no physical principle
         preventing symbol sizes ranging from nanometers or smaller
         to kilometers or larger.  It is possible that
         removal of this  and the other assumptions may require much more
         development of the ideas presented here.

          \end{document}